\documentclass[aps,prb,twocolumn,superscriptaddress,citeautoscript]{revtex4-1}

 \usepackage{svg}
 \usepackage{amsmath,bm}
 \usepackage{mathrsfs}
 \usepackage{amsfonts}
 \usepackage{float}
 \usepackage{setspace} 
 \usepackage{graphicx}
 \usepackage{epstopdf}
 \usepackage{dcolumn}
 \usepackage{amsmath}
 \usepackage{epsfig}
 \usepackage{indentfirst}
 \usepackage{psfrag}
 \usepackage{subfigure}
 \usepackage{amssymb}
 \usepackage{color}
 \usepackage{units} 
\usepackage{physics}
\usepackage{dcolumn}
\usepackage{bm}
\usepackage{bbold}
 \usepackage{dcolumn}
  \usepackage{natbib}
 \usepackage{wrapfig}

\begin{document}


\title{\raggedright On-Surface Synthesis and Characterization of a High-Spin Aza-[5]-Triangulene}

\author{Manuel Vilas-Varela}
  \thanks{Equal Contribution}
  \affiliation{Centro Singular de Investigaci\'on en Qu\'imica Biol\'oxica e Materiais Moleculares (CiQUS) and Departamento de Qu\'imica Org\'anica, Universidad de Santiago de Compostela, 15782 Santiago de Compostela, Spain.}
  
\author{Francisco Romero-Lara}
  \thanks{Equal Contribution}
  \affiliation{CIC nanoGUNE-BRTA, 20018 Donostia-San Sebasti\'an, Spain.}
 
\author{Alessio Vegliante}
  \affiliation{CIC nanoGUNE-BRTA, 20018 Donostia-San Sebasti\'an, Spain.}

\author{Jan Patrick Calupitan}
  \affiliation{Donostia International Physics Center (DIPC), 20018 Donostia-San Sebastian, Spain.}
  \affiliation{Centro de F\'isica de Materiales (CFM-MPC), Centro Mixto CSIC-UPV/EHU, E-20018 Donostia-San Sebasti\'an,  Spain.}

\author{Adri\'an Mart\'inez}  
  \affiliation{Centro Singular de Investigaci\'on en Qu\'imica Biol\'oxica e Materiais Moleculares (CiQUS) and Departamento de Qu\'imica Org\'anica, Universidad de Santiago de Compostela, 15782 Santiago de Compostela, Spain.}

\author{Lorenz Meyer}
  \affiliation{CIC nanoGUNE-BRTA, 20018 Donostia-San Sebasti\'an, Spain.}
   
\author{Unai Uriarte-Amiano}  
  \affiliation{CIC nanoGUNE-BRTA, 20018 Donostia-San Sebasti\'an, Spain.}
  
\author{Niklas Friedrich}  
  \affiliation{CIC nanoGUNE-BRTA, 20018 Donostia-San Sebasti\'an, Spain.}
  
\author{Dongfei Wang}  
  \affiliation{CIC nanoGUNE-BRTA, 20018 Donostia-San Sebasti\'an, Spain.}
 
\author{Natalia E. Koval}  
  \affiliation{CIC nanoGUNE-BRTA, 20018 Donostia-San Sebasti\'an, Spain.}

\author{Mar\'ia E. Sandoval-Salinas}  
  \affiliation{School of Physical and Chemical Sciences, Queen Mary University of London, UK.}

\author{David Casanova}  
  \affiliation{Donostia International Physics Center (DIPC), 20018 Donostia-San Sebastian, Spain.}
  \affiliation{Ikerbasque, Basque Foundation for Science, 48009 Bilbao, Spain.}

\author{Martina Corso}  
  \affiliation{Centro de F\'isica de Materiales (CFM-MPC), Centro Mixto CSIC-UPV/EHU, E-20018 Donostia-San Sebasti\'an,  Spain.}

\author{Emilio Artacho}  
  \affiliation{CIC nanoGUNE-BRTA, 20018 Donostia-San Sebasti\'an, Spain.}
  \affiliation{Theory of Condensed Matter, Cavendish Laboratory, University of Cambridge, J. J. Thomson Ave., Cambridge CB3 0HE, United Kingdom.}
  \affiliation{Ikerbasque, Basque Foundation for Science, 48009 Bilbao, Spain.}
  
\author{Diego Peña}  
  \affiliation{Centro Singular de Investigaci\'on en Qu\'imica Biol\'oxica e Materiais Moleculares (CiQUS) and Departamento de Qu\'imica Org\'anica, Universidad de Santiago de Compostela, 15782 Santiago de Compostela, Spain.}
  \email{diego.pena@usc.es}
  
\author{Jos\'e Ignacio Pascual}
  \affiliation{CIC nanoGUNE-BRTA, 20018 Donostia-San Sebasti\'an, Spain.}
  \affiliation{Ikerbasque, Basque Foundation for Science, 48009 Bilbao, Spain.}
  \email{ji.pascual@nanogune.eu}


\begin{abstract}
    Triangulenes are open-shell triangular graphene flakes with total spin increasing with their size.\ In the last years, on-surface-synthesis strategies have permitted fabricating and engineering triangulenes of various sizes and structures with atomic precision.\ However, direct proof of the increasing total spin with their size remains elusive.\ In this work, we report the combined in-solution and on-surface synthesis of a large nitrogen-doped triangulene (aza-[5]-triangulene) and the detection of its high spin ground state on a Au(111) surface.\ Bond-resolved scanning tunneling microscopy images uncovered radical states distributed along the zigzag edges, which were detected as weak zero-bias resonances in scanning tunneling spectra.\ These spectral features reveal the partial Kondo screening of a high spin state.\ Through a combination of several simulation tools, we find that the observed distribution of radical states is explained by a quintet ground state ($S=2$), instead of the expected quartet state ($S=3/2$), confirming the positively charged state of the molecule on the surface.\ We further provide a qualitative description of the change of (anti)aromaticity introduced by N-substitution, and its role in the charge stabilization on a surface, resulting in a $S=2$ aza-[5]-triangulene on Au(111).
\end{abstract}

\maketitle
\date{\today}

\section*{Introduction}

Triangulenes are triangular-shaped polybenzenoid hydrocarbons with edges formed by \textit{n} zig-zag units (hence, [n]-triangulene) and non-zero electronic spin ground state. The $\pi$-conjugated lattice of [n]-triangulene is frustrated,  depicting a non-Kekulé structure with \textit{n-1} unpaired $\pi$-electrons \cite{Wang09,Su20,deOteyza22,Zeng21,Junzhi20}, forming an electronic ground state with a net spin $S=(n-1)/2$ \cite{Ovchinnikov78}. 
The linear increase of spin state with the triangulene size \textit{n} endows these systems with a strong potential for becoming functional platforms for molecular spintronics and quantum computing applications \cite{Wang09,Han14,Bullard15,Jin18,Sanz22}. 

\begin{figure*}[th!]
    	    \includegraphics[width=\textwidth]{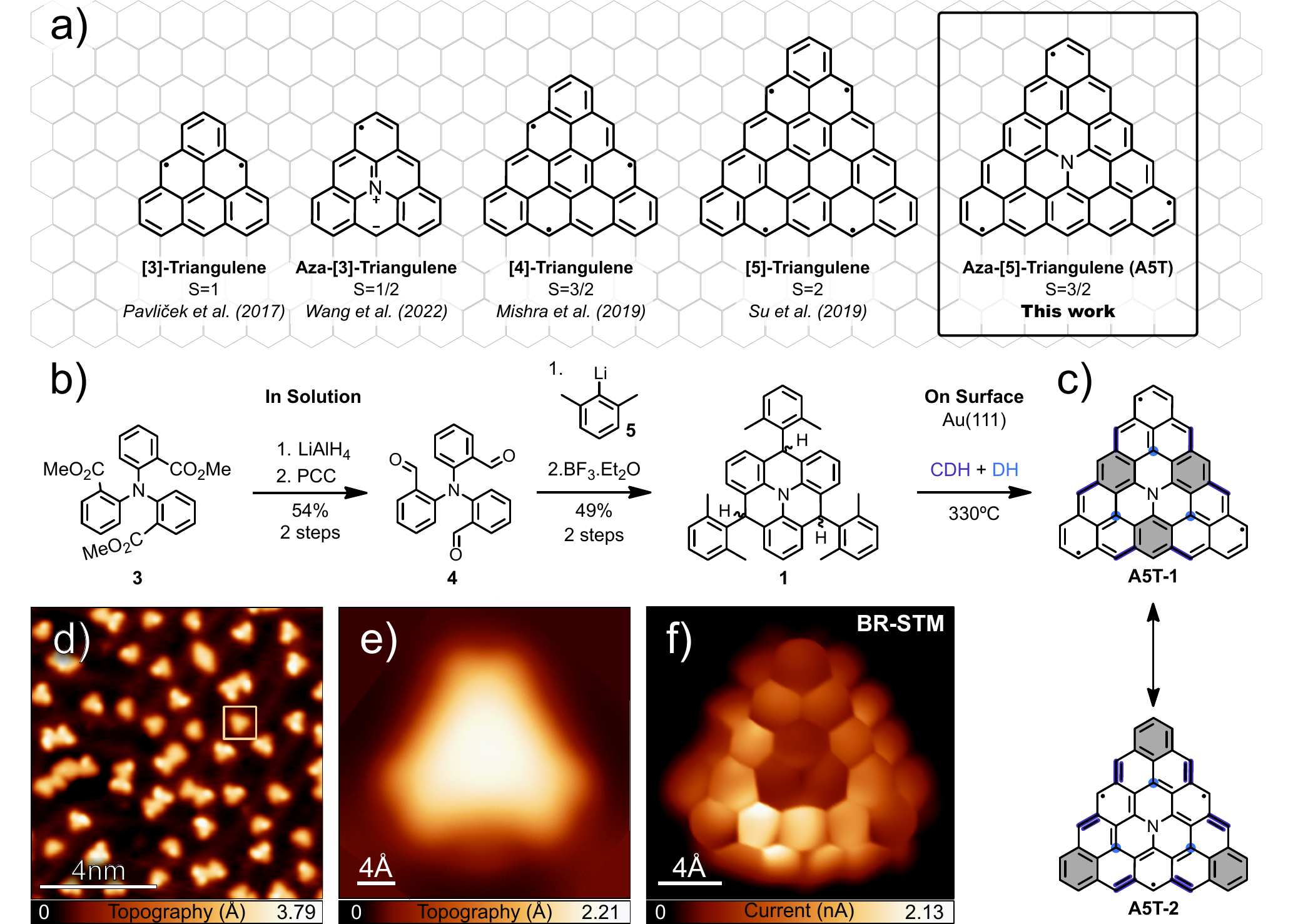}
        	\caption{\label{fig:Fig1} \textbf{a)} Selected previously studied triangulene nanostructures of different sizes and doping. \textbf{b)} Synthetic route to obtain the A5T precursor \textbf{1} by solution chemistry.\ \textbf{c)} Two resonance structures of target A5T showing different locations of Clar sextets (in grey) and $\pi$-radicals. The C-C bonds highlighted in dark blue indicate the ones formed via on-surface assisted cyclodehydrogenation (CDH).\ The light blue spots mark the sites where H atoms were removed via on-surface assisted dehydrogenation (DH).\ \textbf{d)} Resulting STM image ($V=1.5$ V, $I=10$ pA) of molecular precursor \textbf{1} deposited on a Au(111) surface after annealing the sample to 330$^{\circ}$C.\ Yellow square indicates a fully planar structure.\ \textbf{e)} STM image ($V=1$ V, $I=10$ pA) of the A5T. \textbf{f)} Constant height bond resolved STM (BR-STM) image ($V=5$ mV) of the A5T performed with a CO functionalized tip.}
        \end{figure*}

Owing to their open-shell character, the solution synthesis of triangulenes is very challenging \cite{Valenta22,Wei22,Morita11}. Lately, on-surface-synthesis (OSS) strategies  \cite{Cai10,Clair19} have demonstrated to be a viable route for the  fabrication of atomically perfect triangulenes with increasing size \cite{Pavlicek17, Mishra19, Su19,  Mishra21a, Turco23} (some of these shown in Fig.~\ref{fig:Fig1}a). Interestingly, OSS can also produce more complex triangulene nanostructures \cite{Mishra20,Li20a,Mishra20a,Mishra21b,Hieulle21,Su21,Cheng22}, with a variety of magnetic properties emerging from the exchange interaction between triangulene units. 
Overall, the net spin state of triangulene derivatives is associated with an imbalance in the number of carbon sites in the two alternating triangular sublattices ($N_A$ and $N_B$), following the Ovchinnikov's rule \cite{Ovchinnikov78} $S=\frac{1}{2}\abs{N_A-N_B}$.

In addition to tailoring nanographene's shape and size, OSS strategies have also been applied to insert heteroatoms or functional groups for modifying the electronic properties of graphene-based nanostructures \cite{Carbonell-Sanroma17,Wang19,Friedrich20,Li20,Friedrich22,Wen22,Blackwell21,Wang22a}.  Ovchinnikov \cite{Ovchinnikov78} predicted that heteroatom substitution in alternant sublattices acts as a defect that modifies the sublattice imbalance and, hence, the resulting spin state \cite{Ugeda10,Gonzalez-Herrero16}. However, recent results on aza-[3]-triangulene (A3T) \cite{Sandoval-Salinas19,Wang22a,Wei22} found that the nitrogen substitution in minority sites reduces the spin state from $S=1$ in all-carbon [3]-triangulene (3T) to $S=1/2$ in A3T. This is in apparent contradiction with Ovchinnikov's prediction that a larger spin ground state ($S=3/2$ in A3T) shall be expected when minority sites are removed. Investigation of this apparent paradox requires experimental access to larger triangulene structures, and determination of their spin state.   

Despite the successful imaging of triangulenes on metal surfaces using low-temperature scanning tunneling microscopy (STM) \cite{Turco23,Wang22a,Mishra19,Su19,Mishra21a,Chen22}, the resolution of their intrinsic spin state has been hampered by the lack of spin-sensitive signals. Owing to the very weak magnetic anisotropy of carbon systems, the $\pi$-magnetism is fairly isotropic and paramagnetic, thus difficult to access by spin-polarized tunneling microscopy \cite{Brede23}. Instead, a zero-bias resonance due to the Kondo-screening \cite{Kondo64,Ternes09} of spins by the metallic substrate has been normally used as an unequivocal fingerprint of a spin-polarized ground state in triangulenes \cite{Li19,Wang22a,Turco23}. Unfortunately, the universal behavior of Kondo screening determines that the associated zero-bias resonance decreases its intensity with increasing spin values \cite{Parks10,Li20a,Wang22a}, requiring very low temperatures for its detection. 
 
In this work, we present an OSS strategy for engineering a large nitrogen-doped triangulene, the aza-[5]-triangulene (A5T, rectangle in Fig.~\ref{fig:Fig1}a), and demonstrate that it lies in an S=2 ground state on a Au(111) surface. The synthesis route involved the targeted thermal cyclodehydrogenation of a trisubstituted A3T derivative (\textbf{1} in Fig.~\ref{fig:Fig1}b) over the gold substrate.\ Combining bond-resolved STM and spectral maps of the density of states close to the Fermi level, we show that A5T lies in a high-spin ground state, comprised of four singly-occupied states. Our density functional theory (DFT) simulations reveal that the nitrogen heteroatom, located in the majority sublattice in A5T, reduces the S=2 spin of the pristine [5]-triangulene (5T) down to 3/2.\ Although this coincides with the prediction by Ovchinnikov's rule, the origin is caused by the addition of an extra electron in the $\pi$-conjugated system, donated by the N atom, which also induces a marked anti-aromatic character around the aza group. On the metal substrate, the A5T system transfers this extra electron to the metal underneath, i.e., it is oxidized to A5T$^+$, recovering the quintet ground state of the parent 5T flake.\ These results validate predictions for high-spin nanographenes \cite{Wang07,Su19} and further demonstrate that such a high-spin state survives over a metallic substrate.
 
   \begin{figure*}[th!]
    	\includegraphics[width=\textwidth]{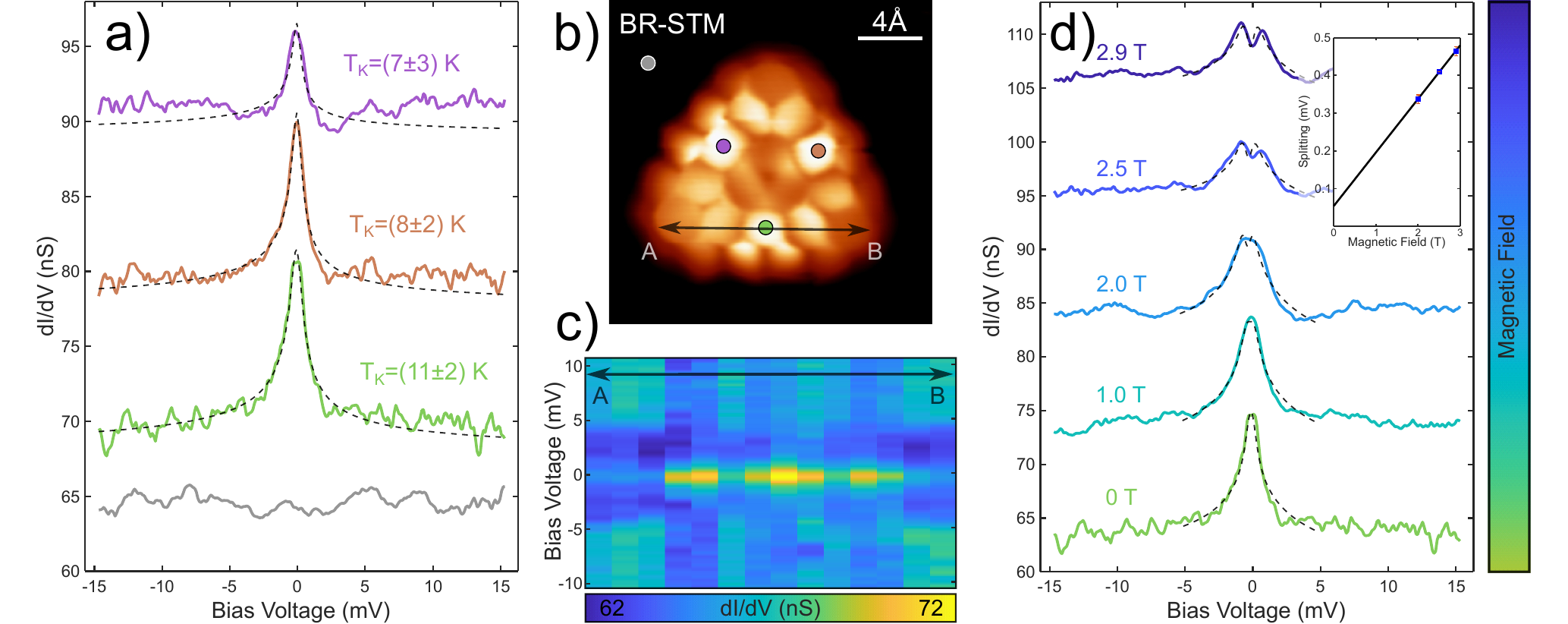}
            \caption{\label{fig:Fig2}
            \textbf{a)} Low-energy dI/dV spectra taken on the spots marked by colored circles in b) at $T=1.2$ K.\ Dashed lines are fits of the three spectra using a Frota function \cite{Frota92}, from which the indicated Kondo temperatures ($T_K=\sqrt{\textrm{FWHM}^2-(2\pi k_BT)^2}/2k_B$ \cite{Gruber18}) are estimated assuming a strong-coupling Kondo regime ($T<T_K$). Spectra are vertically shifted for clarity. 
            \textbf{b)} BR-STM image ($V=5$ mV) of A5T.\
            \textbf{c)} Low-energy dI/dV spectral line along one edge of A5T, marked by an arrow in b), obtained with a CO-functionalized tip.\
            \textbf{d)} Magnetic field-dependent low-energy dI/dV measurements.\ Spectra were taken on the green spot marked in b). Dashed lines correspond to the fits of the Kondo resonances in a magnetic field for a spin-2 model including third-order terms with the code from Ternes \cite{Ternes15}. Spectra were shifted vertically for clarity. The inset shows the average splitting of the Kondo peak fits as a function of the magnetic field for three sets of measurements (on each of the  spots of the A5T in image b)). The splittings follow the line $E_{\mathrm{ZS}}=g\mu_BSB$ with a fitted $g=1.23\pm0.12$. More details about the procedure are in Fig. S4.}
             
        \end{figure*}

\section*{Results and Discussion}

\textbf{Solution and on-surface synthesis of A5T:} 
Inspired by the precursor design used by Su et al. for the generation of 5T \cite{Su19}, we envisioned the synthesis of A5T by on-surface planarization of the A3T derivative \textbf{1}, which is substituted with three dimethylphenyl groups (Fig.~\ref{fig:Fig1}b). Compound \textbf{1} was obtained by solution chemistry in four steps from amine \textbf{3} \cite{Wang22a}. First, sequential treatment with lithium aluminum hydride (LiAlH$_4$), followed by oxidation with pyridinium chlorochromate (PCC) afforded trialdehyde \textbf{4} in 54\% yield. Then, the addition of three equivalents of organolithium derivative \textbf{5}, followed by BF$_3$-promoted three-fold intramo-lecular Friedel–Crafts, led to the formation of the A5T precursor \textbf{1} in 49\% yield.

Precursor \textbf{1} was deposited on a Au(111) substrate at room temperature and in ultra-high vacuum conditions. To obtain the targeted A5T, six new C-C bonds had to be created, and fifteen hydrogen atoms had to be removed from its bulky three-dimensional structure. We annealed the pre-covered substrate to 330 $^\circ$C (see methods) to induce cyclodehydrogenation (CDH) and dehydrogenation (DH) of the precursor and obtained a sample with planar molecular platforms, as we confirmed using low-temperature STM. Images like in Fig.~\ref{fig:Fig1}d resolve that some molecules appeared with some bright lobes corresponding to sp3 carbon atoms surviving the CDH/DH step, while others underwent covalent couplings between flakes forming larger structures. These partially reacted molecules tend to self-assemble and $\pi$-stack and can be planarized by tip manipulation (see Fig. S1).\ Nevertheless, fully planarized structures corresponding to the target A5T molecule, (square in Fig.~\ref{fig:Fig1}d, zoomed in Fig.~\ref{fig:Fig1}e) can be found, allowing an in-depth study of their electronic and magnetic structure. 

Figure~\ref{fig:Fig1}f shows a high-resolution STM image of the intact A5T flake from Fig.~\ref{fig:Fig1}e, obtained with a CO-functionalized tip at a low bias (typically $\leq5$ mV) and scanning in cons-tant-height mode with a tunneling resistance of around 20 M$\Omega$ \cite{Gross09,Kichin11}. This allows us to achieve a bond-resolved (BR) image that shows the honeycomb lattice of the A5T and confirms the successful synthesis. In some cases, remnant H-atoms from incomplete CDH show up in the BR images as deformations of the zigzag edges (e.g., as in Fig. S2). However, they were removed by high-energy electron tunneling \cite{Li19}, until the final product, A5T, was obtained, like in Fig.~\ref{fig:Fig1}f. 
Interestingly, the current contrast in the BR images is not homogeneous across the A5T backbone but appears higher along zigzag edges and darker at the center and at the three vertexes. The larger current over the edges suggests the presence of zero-energy features, normally related to spin states. The lower tunneling current over the A5T center is probably due to both the absence of zero-energy features and a small out-of-plane distortion of the N towards the Au(111) surface. The darker rings at the vertexes also claim spin-free regions, such as one would expect if highly stable Clar sextets \cite{Clar83,Sola13} were localized at these sites. Tentatively, one may thus compare this tunneling current distribution with the dominant resonant structures expected for A5T (Fig.~\ref{fig:Fig1}c), and find that  structure \textbf{A5T-2} appears closer to the experimental tunneling current maps. 

        \begin{figure*}[t!]
            \includegraphics[width=\textwidth]{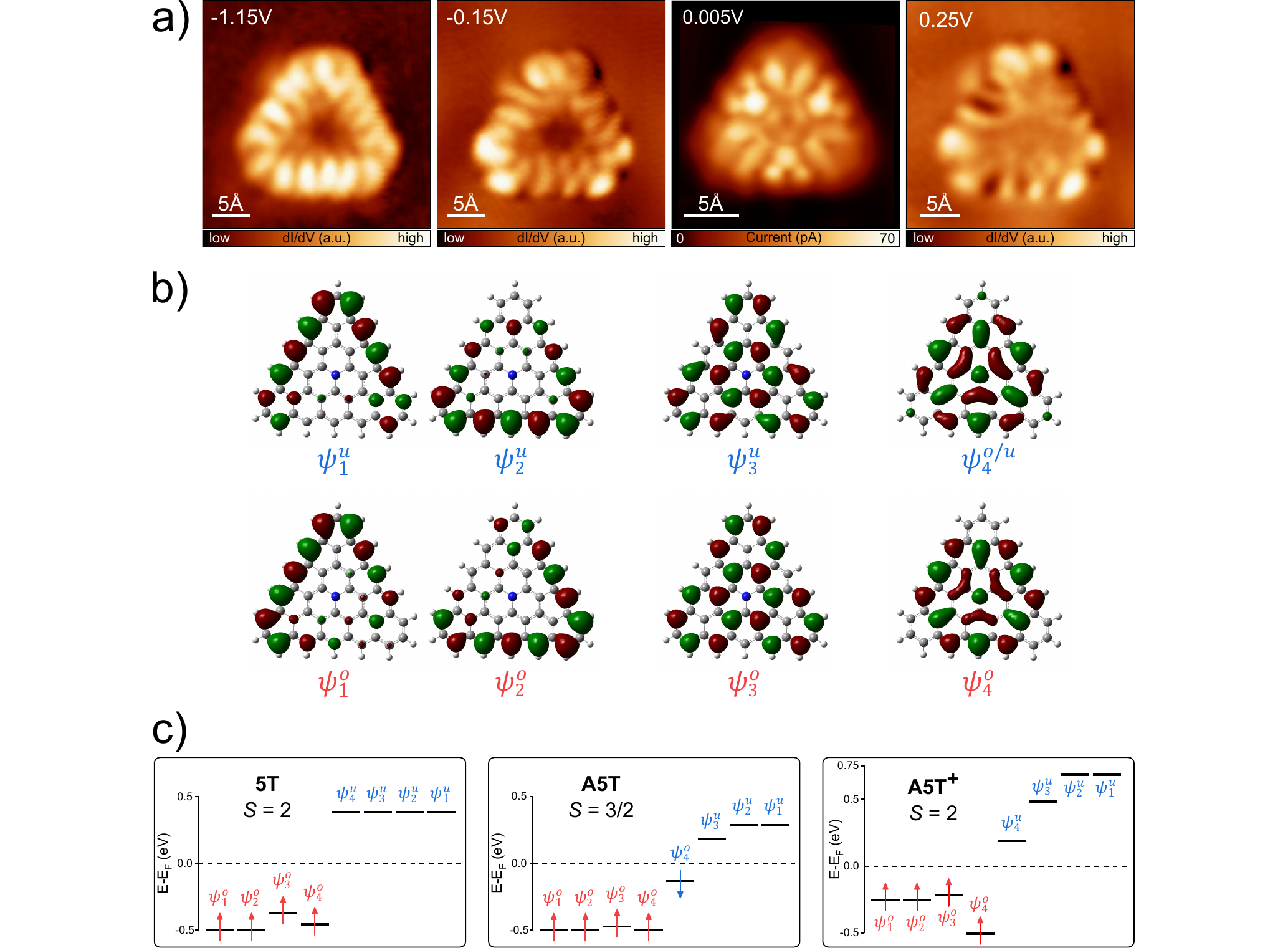}
            \caption{\label{fig:Fig3} \textbf{a)} dI/dV maps at different bias acquired at constant current ($I=500$ pA) of A5T. The one corresponding to 0.005 V maps the orbital that contributes the most to the Kondo resonance. In this case, instead of the dI/dV signal, the current is recorded at constant height with a CO-functionalized tip at larger distances from the molecule as compared to the BR images. \textbf{b)}  Molecular orbital isosurfaces of the A5T and the A5T$^+$ of the SOMO-SUMO obtained from spin-polarized DFT. \textbf{c)} Energy levels diagram obtained from spin-polarized DFT calculations close to Fermi energy for the case of 5T, A5T, and A5T$^+$.}
        \end{figure*}

\textbf{Analysis of the Kondo effect to reveal a high spin state:}     
In order to experimentally address the spin ground state of A5T, we performed tunneling spectroscopy at $T=1.2$ K, measuring the low-bias differential conductance signal (dI/dV) over various parts of the molecule. Fig.~\ref{fig:Fig2}a shows $dI/dV$ vs. bias spectra taken on the central rings of the three edges of the A5T molecule (colored dots in Fig.~\ref{fig:Fig2}b). Narrow zero-bias spectral peaks reveal the existence of a Kondo-screened spin state at these sites. To capture the spatial distribution of the spin signal, we measured a spectral profile along the edges of the A5T molecule (Fig.~\ref{fig:Fig2}c).\ The Kondo resonance extends along the three central rings with its maximum intensity above the central one but decreases towards the vertexes, as well as over the central region of the molecule as in Fig. S3.\ This distribution of the Kondo signal agrees with the tunneling current variations of the BR-images. The map of Fig.~\ref{fig:Fig2}c also reveals that the amplitude of the Kondo peak is very small, only slightly larger than the small inelastic steps at $\pm5$ mV (also in Fig. S3) attributed to the excitation of the frustrated translational vibrational mode of the CO molecule attached to the tip apex \cite{delaTorre17}.

The narrow line-width (FWHM $<$ 2 mV) of the Kondo resonances are consistent with a Kondo temperature $\sim 10$ K (Fig.~\ref{fig:Fig2}a), larger than the temperature of the measurement, indicating that the A5T molecules lie in the strong-coupling Kondo regime. However, zero-bias peaks with such a small amplitude are normally observed when the molecular system has a spin larger than $S=1/2$, and the Kondo effect at the measuring temperature only applies to some of the spin channels, i.e., an underscreened Kondo effect \cite{Li20a}. The underscreened Kondo character of A5T can be further demonstrated by measuring the evolution of the zero-bias resonance under increasing magnetic fields.   Fig.~\ref{fig:Fig2}d shows the spectra measured on the green spot marked in Fig.~\ref{fig:Fig2}b as the magnetic field is ramped up to $2.9$ T. The Kondo resonance splits already for low values of the magnetic field, instead of slowly broadening with increasing magnetic field, as expected for a fully screened case \cite{Li19}. As shown in the inset of Fig.~\ref{fig:Fig2}d, the splitting energy follows the Zeeman energy of an $S=1/2$ in a magnetic field, indicating that the Kondo screening is not complete \cite{Li20a}. Hence, the Kondo feature suggests a large spin ground state for A5T, but, we cannot obtain a precise indication of its total spin from its magnetic field dependence. 
This is because, in the absence of magnetic anisotropy, the split of the Kondo resonance always follows $g\mu_b$, the $S=1/2$ Zeeman excitation energy  \cite{Parks10}.

    \begin{figure*}[t!]
            \includegraphics[width=\textwidth]{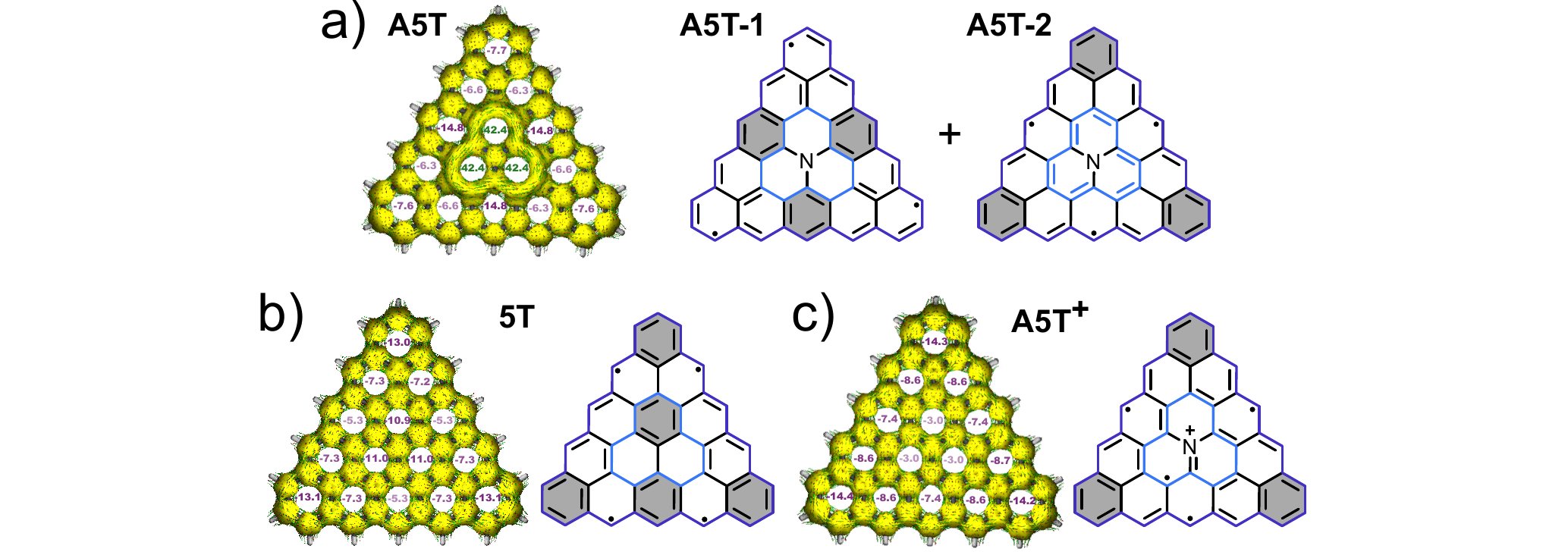}
            \caption{\label{fig:Fig4} Map of the computed anisotropy of the induced current density (ACID), where the numbers inside the benzene rings indicate the nucleus-independent chemical shift $ZZ$ indexes 1 \r{A} above the molecular plane (NICS(1)$_{zz}$), and resonance forms for the case of \textbf{a)} A5T, \textbf{b)} 5T, and \textbf{c)} A5T$^+$. The C-C bonds in dark/light blue  correspond to the outer/inner rims.}
        \end{figure*}

\textbf{Identification of frontier states:} 
To find out the spin state, we analyze the orbital structure around the Fermi energy (E$_F$) of the system using dI/dV maps and compare them to results from DFT and mean-field Hubbard (MFH) simulations (see methods). Fig.~\ref{fig:Fig3}a shows a set of selected differential conductance maps at different bias values around zero bias, i.e. Fermi level (see Fig. S5 for additional bias values in the range of (-1.85, 2.05) V). Albeit dI/dV plots do not show marked resonances, the dI/dV maps exhibit clear shapes, corresponding to the local density of states of the different molecular orbitals of A5T. Fig.~\ref{fig:Fig3}b and 3c show the wavefunctions and the energy level diagram respectively of the spin-unrestricted orbitals of A5T computed by DFT (see methods) around Fermi level ($\pm 0.75$ eV). The $C_{3v}$ symmetry of the molecule on the surface imposes degeneracy into $\psi_1$ and $\psi_2$ orbitals while the two others remain non-degenerate. For the case of A5T (mid panel in Fig.~\ref{fig:Fig3}c), $\psi_1^{o(u)}$, $\psi_2^{o(u)}$, and $\psi_3^{o(u)}$ corresponds to the singly (un)occupied molecular orbitals SOMO(SUMO), while $\psi_4$ remains fully occupied. This results in a $S=3/2$ spin ground state, also in good agreement with MFH calculations in Fig. S6.

To reveal the spin state on the surface, we compare the orbital distribution of the computed SOMOs with the dI/dV maps of Fig.~\ref{fig:Fig3}a. The most important feature here is that the zero-bias map, which reproduces  the amplitude distribution of the Kondo resonance, follows the characteristic shape of the  $\psi_4$ orbital, accounting for a larger signal on the edges than on the vertexes. This indicates that $\psi_4$ is the Kondo-screened state, evidencing its singly occupied character on the Au(111) substrate, as it would be in the case of the cationic species A5T$^+$ in the right panel of Fig.~\ref{fig:Fig3}c. The rest of the SOMOs can be also recognized in the dI/dV maps of Fig.~\ref{fig:Fig3}a ($\psi^o_{1,2}$ at $\sim$-1.15 eV, and the correlated pair $\psi^o_3$ and $\psi^u_3$ at $\sim$ -0.15 eV and 0.25 eV, respectively). In other words, the $S=2$ ground state is a consequence of the charge donation from $\psi_4^{o/u}$ to the surface resulting in a positively charged molecule, as similarly observed in A3T \cite{Wang09}.

The effect of the N heteroatom substitution can be deduced by comparing its orbital structure with that of the all-carbon [5]-triangulene (5T) \cite{Su19}. Owing to its particular topology, 5T has four SOMOs, i.e., a nullity $\eta$=4 \cite{Wang09}. This gives rise to a $S=2$ ground state in the presence of Coulomb correlations, as confirmed by DFT (Fig.~\ref{fig:Fig3}c and Fig. S7) and MFH (Fig. S6). Substituting the central carbon atom with a nitrogen atom does not distort the orbital shapes (see orbital shapes of 5T in Fig. S6 and S7), but simply adds an extra electron into the $\pi$-conjugated network. This ``extra" $\pi$-electron populates the $\psi_4$ state, which is the state with the largest amplitude at the N site (Fig.~\ref{fig:Fig3}b and Fig. S7), and, consequently reduces the spin to $S=3/2$  (from left to mid panel in Fig.~\ref{fig:Fig3}c).  Contrary to the smaller A3T, the extra electron does not lead to Jahn-Teller distortions \cite{Sandoval-Salinas19}, because here it populates a non-degenerate level. Consequently, the lack of stabilizing Jahn-Teller distortion keeps the $\psi_4$ state closer to the Fermi level than the other frontier orbitals. Therefore, the electron added by the N heteroatom to the conjugated $\pi$-system remains the most valent and is donated to the Au(111) substrate, eventually leading to the oxidized A5T (A5T$^+$), which adopts the $S=2$ ground state of 5T on the surface.

\textbf{Global and local (anti)aromaticity:} Spectral maps of Kondo-amplitude like in Fig.~\ref{fig:Fig1}f, ~\ref{fig:Fig2}b and ~\ref{fig:Fig3}a resolved a peculiar pattern of unpaired electron localization at the zigzag edges of the flakes. In fact, these maps resemble a  resonant structure like \textbf{A5T-2} in Fig.~\ref{fig:Fig4}a, with radical states delocalized within the middle zigzag edges, rather than the \textbf{A5T-1} structure, where radicals lie at the triangulene vertexes. Here, we  theoretically analyze the (anti)aromaticity of the A5T molecule and show that such an electron delocalization pattern is a fingerprint for the spin state of the flake on the surface.

In Fig.~\ref{fig:Fig4}a we plot the  anisotropy of the induced current density (ACID)\cite{Geuenich05} and the nucleus independent chemical shift \cite{Schleyer96} (NICS indexes in ppm included in the ACID maps of Fig.~\ref{fig:Fig4}) computed for the A5T molecule in the gas phase.\ The A5T molecule has a peripheral triangular rim formed by 30 C-atoms (dark blue C-C bonds in Fig.~\ref{fig:Fig4}), which would fulfill the 4n+2 H\"uckel rule for aromaticity if each atom contributed with a $\pi$-electron. However, the corresponding ACID plot for a neutral A5T  finds a very small ring current around this outer 30-carbon circuit, which is a sign of the global non-aromatic nature of the outer rim. This points to the presence of unpaired electrons at the edges or vertexes, reducing the H\"uckel counting rule, as described by both structures \textbf{A5T-1} and \textbf{A5T-2} in Fig.~\ref{fig:Fig4}a. 
In fact, NICS indexes find that the benzene rings at the periphery present some (local) aromatic character (NICS(1)$_{zz}<0$ ppm), with larger negative NICS at the middle-edge rings, and to some extent also at the rings at the vertexes. This indicates a larger probability for hosting Clar sextets at these sites \cite{Portella05}, as in \textbf{A5T-1} structure.  
However, the ACID plot of neutral A5T also displays an anticlockwise paramagnetic current flow in the inner rim (light blue C-C bonds in Fig.~\ref{fig:Fig4}), which can be traced to the [12]annulene moiety present in the \textbf{A5T-2} structure, expected to exhibit global antiaromaticity. This is confirmed by the large NICS positive values (NICS(1)$_{zz}$ = 42.4 ppm) of the three central rings.  Based on this electron delocalization and aromaticity indices we conclude that the neutral A5T molecule lies in a superposition of both, the non-aromatic \textbf{A5T-1} and the antiaromatic \textbf{A5T-2} resonant structures.

The antiaromatic character of the inner rim of A5T contrasts with their aromatic character in the all-carbon 5T flake (Fig.~\ref{fig:Fig4}b). Our simulations (both DFT and MFH) suggest that the local antiaromaticity over the center is brought by the extra $\pi$ electron inserted by the aza group, because it doubly populates the state with larger orbital amplitude around the center ($\psi^4$), and forces the formation of the [12]annulene inner rim. The aza-moiety thus endows the flake with an intrinsic tendency to donate charge by inserting a hole into the doubly occupied $\psi^4$ state because this reduces the antiaromaticity of the center, a process that is favored by substrates with high electron affinity, such as Au(111).

The A5T$^+$ cation is built up by a hole extending along the $\psi^4$ orbital (Fig.~\ref{fig:Fig3}b), which is the Kondo screening channel.
In fact, the $\psi^4$-hole contributes to breaking the conjugation of the [12]annulene circuit.\ The corresponding ACID plot in Fig.~\ref{fig:Fig4}c shows no paratropic currents, and a pattern very similar to the all-carbon 5T flake (Fig.~\ref{fig:Fig4}b) appears. The transition from antiaromatic to non-aromatic character is confirmed by NICS(1)$_{zz}$ values computed for A5T$^+$ (Fig.~\ref{fig:Fig4}c). The  three inner rings become non-aromatic in the cationic flake, and the vertexes host now the most aromatic rings (Clar sextet). 
Therefore, we conclude that the preferential spin localization observed in the experimental Kondo map of Fig.~\ref{fig:Fig3}a (also hinted from the BR-STM images of Figs.~\ref{fig:Fig1}f and ~\ref{fig:Fig2}b), reproduces the features of the cationic character of the A5T$^+$ cation and, hence, confirms its $S=2$ ground state.

\section*{Conclusion}

In summary, we have described a route to fabricate a large aza-triangulene flake on a metal substrate and demonstrated that it lies in a high spin state.\ The fabrication was realized by combining solution synthesis of rationally-designed molecular precursors, and on-surface synthesis by thermally-induced dehydrogenation reactions. We demonstrated the magnetic state by detecting a very weak Kondo resonance originating from a molecular state with a hole radical caused by charge donation to the surface. Combining experimental orbital maps, Kondo amplitude maps, DFT and MFH simulations, and calculated ACID plots and NICS indexes, we determined that the aza-triangulene flake lies in the cationic $S=2$ state on the Au(111) substrate, thus representing a high-spin nanographene. Our work confirms that aza-triangulenes are prone to act as charge donors, as a way to increase their aromatic character.

\section*{Acknowledgements}

The authors gratefully acknowledge financial support from grants No. PID2019-107338RB, PID2019-109555GB-I00, FIS2017-83780-P, and CEX2020-001038-M funded by MCIN / AEI / 10.13039 / 501100011033, the ELKARTEK project BRTA QUANTUM (no. KK-2022/00041), the European Regional Development Fund, and the European Union (EU) H2020 program through the FET Open project SPRING (grant agreement No.~863098). F.R.-L. thanks the Spanish Ministerio de Educaci\'{o}n y Formaci\'{o}n Profesional through the PhD scolarship no. FPU20/03305. M.E.S.-S. acknowledges the funding by the UK Research and Innovation under the UK government’s Horizon Europe funding guarantee (grant number EP/X020908/1). We thank Thomas Frederiksen, Sofía Sanz, and Ricardo Ortiz for fruitful discussions.

 
%

\end{document}